\documentclass[journal]{IEEEtran}
  \def\fH{{\cal H}}

\usepackage{theorem}
\theorembodyfont{\rmfamily}
\newtheorem{Lem}{Lemma}[section]
\newtheorem{Def}[Lem]{Definition}
\newtheorem{The}[Lem]{Theorem}
\newtheorem{Prop}[Lem]{Proposition}

\newtheorem{Rem}[Lem]{Remark}

\newcommand{\qed}{\hbox{\rule{6pt}{6pt}}}

\begin{document}
%\usepackage{amssymb}
%\usepackage{latexsym}
% paper title
\title{A generalized skew information and uncertainty relation}

\author{Kenjiro~Yanagi,~\IEEEmembership{Member,~IEEE,}
        Shigeru~Furuichi,~\IEEEmembership{Member,~IEEE,}
        and~Ken~Kuriyama,~\IEEEmembership{}% <-this % stops a space
\thanks{Manuscript received ; revised .}% <-this % stops a space
\thanks{K.Yanagi is with the Department of Applied Science, Yamaguchi University, Ube City, Yamaguchi, 755-8611, Japan, Email: yanagi@yamaguchi-u.ac.jp.}
\thanks{S.Furuichi is with Department of Electronics and Computer Science,
Tokyo Univ. of Science in Yamaguchi, SanyoOnoda City, 756-0884, Japan, Email: furuichi@ed.yama.tus.ac.jp.}
\thanks{K.Kuriyama is with the Department of Applied Science, Yamaguchi University, Ube City, Yamaguchi, 755-8611, Japan, Email: kuriyama@yamaguchi-u.ac.jp.}
}

\maketitle

\begin{abstract}%{\bf Abstract.} 
A generalized skew information is defined and a generalized uncertainty relation is established with the help of a trace inequality
 which was recently proven by J.I.Fujii. In addition, we prove the trace inequality conjectured by S.Luo and Z.Zhang.
Finally we point out that Theorem 1 in {\it S.Luo and Q.Zhang, IEEE Trans.IT, Vol.50, pp.1778-1782 (2004)} is incorrect in general, by giving a simple counter-example.
\end{abstract}
%\vspace{3mm}

%{\bf Mathematics Subject Classification 2000:} 15A42, 47A63 and 94A17
%\vspace{3mm}

%\vspace{3mm}
\IEEEpeerreviewmaketitle
%%%%%%%%%%%%%%%%%%%%%%%%%%%%%%%%%%% Section %%%%%%%%%%%%%%%%%%%%%%%%%%%%%%%%%%%%%%%%%%%%%%%%%%%%%%%%%%%

\section{Introduction} \label{sec1}

%Two important theorems \cite{Schu,Hol} on quantum information theory respectively showed the relations between (1) quantum entropy \cite{Neu} and source coding theorem in quantum system, 
%(2) Holevo bound and coding theorem for the classical-quantum channel.
%In \cite{Schu}, the role of von Neumann entropy \cite{Neu} in quantum information was clarified so that quantum information theory
%has been progressed for a decade \cite{Hol,NC}.  Before such developments of quantum information, von Neumann entropy and related entropies such as
% relative entropy \cite{Ume} and mutual entropy \cite{Ohy} were studied in both direction from physics and mathematics \cite{Weh,OP}. 

As one of the mathematical studies on entropy, the skew entropy \cite{WY1,WY2} and the problem of its concavity are famous. The concavity
problem for the skew entropy generalized by F.J.Dyson, was solved by E.H.Lieb in \cite{Lie}. It is also known that the skew entropy
represents the degree of noncommutativity between a certain quantum state represented by the density matrix $\rho$
 (which is a positive semidefinite matrix with unit trace) and an observable represented by the selfadjoint matrix $X$. 
Quite recently S.Luo and Q.Zhang studied the relation between skew information 
(which is equal to the opposite signed skew entropy) and the uncertainty relation in \cite{LZ}.
Inspired by their interesting work, we define a generalized skew information and then study the relationship between it and the uncertainty relation.
In addition, we prove the trace inequality conjectured by S.Luo and Z.Zhang in \cite{LZ2}.

%%%%%%%%%%%%%%%%%%%%%%%%%%%%%%%%%%% Section %%%%%%%%%%%%%%%%%%%%%%%%%%%%%%%%%%%%%%%%%%%%%%%%%%%%%%%%%%%
\section{Preliminaries}

Let $f$ and $g$ be functions on the domain $D \subset \hbox{R} $.
$(f,g)$ is called a monotonic pair if $(f(a) - f(b)) (g(a) - g(b)) \geq 0$ 
for all $a,b \in D$. $(f,g)$ is also called an antimonotonic pair if 
$(f(a) - f(b)) (g(a) - g(b)) \leq 0$ for all $a,b \in D$. 

In what follows we consider selfadjoint matrices whose spectra are included in $D$ so that functional calculus makes sense.
\begin{Lem} (\cite{Bou,Fuj}) \label{Fujii}
For any selfadjoint matrices $A$ and $X$, we have the following trace inequalities.
\begin{itemize}
\item[(1)] If $(f,g)$ is a monotonic pair, then 
$$\hbox{Tr}\left(f(A)Xg(A)X\right) \leq \hbox{Tr}\left(f(A)g(A)X^2\right). $$
\item[(2)] If $(f,g)$ is an antimonotonic pair, then 
$$\hbox{Tr}\left(f(A)Xg(A)X\right) \geq \hbox{Tr}\left(f(A)g(A)X^2\right). $$
\end{itemize}
\end{Lem}
From this lemma, we can obtain the following lemma.

\begin{Lem}   \label{g_Fujii}
For any selfadjoint matrices $A$ and $B$, and any matrix $X$, we have the following trace inequalities.

\begin{itemize}
\item[(1)] If $(f,g)$ is a monotonic pair, then 
\begin{eqnarray*}
&& \hbox{Tr}\left(f(A)X^*g(B)X + f(B)Xg(A)X^*\right)\\
&& \leq \hbox{Tr}\left(f(A)g(A)X^*X +f(B)g(B)XX^*\right). 
\end{eqnarray*}
\item[(2)] If $(f,g)$ is an antimonotonic pair, then 
\begin{eqnarray*}
&& \hbox{Tr}\left(f(A)X^*g(B)X + f(B)Xg(A)X^*\right)\\
&& \geq \hbox{Tr}\left(f(A)g(A)X^*X +f(B)g(B)XX^*\right). 
\end{eqnarray*}
\end{itemize}
\end{Lem}
{\it Proof} :
Define on $\fH  \oplus  \fH$ 
\[
\widehat{A} = \left( \begin{array}{l}
 A\,\,\,\,0 \\ 
 0\,\,\,\,B \\ 
 \end{array} \right),\widehat{X} = \left( \begin{array}{l}
 \,0\,\,\,\,X^*  \\ 
 X\,\,\,\,0 \\ 
 \end{array} \right),
\]
where $A,B$ and $X$ act on a finite dimensional Hilbert space $\fH$.
Then $\widehat{A}$ and $\widehat{X}$ are selfadjoint. Therefore one may apply Lemma \ref{Fujii} to get 
\begin{eqnarray*}
&&\hbox{Tr}\left(f(A)X^*g(B)X+f(B)Xg(A)X^*\right)\\
&&=\hbox{Tr}\left( \left( \begin{array}{l}
 f\left( A \right)\,\,\,\,\,\,\,\,0 \\ 
 \,\,\,\,0\,\,\,\,\,\,\,\,\,f\left( B \right) \\ 
 \end{array} \right)\left( \begin{array}{l}
 \,0\,\,\,\,X^*  \\ 
 X\,\,\,\,0 \\ 
 \end{array} \right) \right. \\
&&\left.\left( \begin{array}{l}
 g\left( A \right)\,\,\,\,\,\,\,\,0 \\ 
 \,\,\,\,0\,\,\,\,\,\,\,\,\,g\left( B \right) \\ 
 \end{array} \right)\left( \begin{array}{l}
 \,0\,\,\,\,X^*  \\ 
 X\,\,\,\,0 \\ 
 \end{array} \right) \right) \\ 
&&=\hbox{Tr}\left(f(\widehat{A})\widehat{X}g(\widehat{A})\widehat{X}\right)\\
&&\leq \hbox{Tr}\left(f(\widehat{A})g(\widehat{A})\widehat{X}^2\right)\\
&&=\hbox{Tr}\left( \left( \begin{array}{l}
 f\left( A \right)\,\,\,\,\,\,\,\,0 \\ 
 \,\,\,\,0\,\,\,\,\,\,\,\,\,f\left( B \right) \\ 
 \end{array} \right)\left( \begin{array}{l}
 g\left( A \right)\,\,\,\,\,\,\,\,0 \\ 
 \,\,\,\,0\,\,\,\,\,\,\,\,\,g\left( B \right) \\ 
 \end{array} \right) \right. \\
 &&\left. \left( \begin{array}{l}
 \,0\,\,\,\,X^*  \\ 
 X\,\,\,\,0 \\ 
 \end{array} \right)\left( \begin{array}{l}
 \,0\,\,\,\,X^*  \\ 
 X\,\,\,\,0 \\ 
 \end{array} \right) \right) \\ 
&&=\hbox{Tr}\left(f(A)g(A)X^*X+f(B)g(B)XX^*\right),
\end{eqnarray*}
which is inequality (1).  Inequality (2) is proven in a similar way.
\hfill \qed

%%%%%%%%%%%%%%%%%%%%%%%%%%%%%%%%%%% Section %%%%%%%%%%%%%%%%%%%%%%%%%%%%%%%%%%%%%%%%%%%%%%%%%%%%%%%%%%%

\section{Generalized uncertainty relation}

For a density matrix (quantum state) $\rho$ and arbitrary matrices $X$ and $Y$ acting on $\fH$, we denote
$\widetilde{X} \equiv X - \hbox{Tr}\left(\rho X\right)I$ and $\widetilde{Y} \equiv Y - \hbox{Tr}\left(\rho Y\right)I$, 
where $I$ represents the identity matrix. Then
we define the covariance by $\hbox{Cov}_{\rho}(X,Y) = \hbox{Tr}\left(\rho \widetilde{X} \widetilde{Y}\right)$. Each variance is defined by
$V_{\rho}(X) \equiv \hbox{Cov}_{\rho}(X,X)$ and $V_{\rho}(Y) \equiv \hbox{Cov}_{\rho}(Y,Y)$. 

The famous Heisenberg's uncertainty relation \cite{Hei,Rob} can be easily proven by the application of the
Schwarz inequality and it was generalized by Schr\"odinger as follows:

\begin{Prop} (Schr\"odinger \cite{Schr})  \label{ur}
For any density matrix $\rho$ and any two selfadjoint matrices $A$ and $B$, we have the uncertainty relation :
\begin{equation} \label{eq_ur}
V_{\rho}(A) V_{\rho}(B) - \vert \hbox{Re}\left( \hbox{Cov}_{\rho}(A,B) \right)\vert ^2 \geq \frac{1}{4} \vert \hbox{Tr}\left(\rho[A,B]\right)\vert^2, 
\end{equation}
where $[X,Y] \equiv XY-YX$.
\end{Prop}
\begin{Def}
For arbitrary matrices $X$ and $Y$, we define
$$
I_p(\rho;X,Y) \equiv \hbox{Tr}\left(\rho XY\right) - \hbox{Tr}\left(\rho^{\frac{1}{p}}X\rho^{\frac{1}{p^*}}Y\right),
$$
where $p \in [1,+\infty]$ and with $p^*$ such that $\frac{1}{p} + \frac{1}{p^*} = 1$.
If $A$ is selfadjoint, the Wigner-Yanase-Dyson information is defined by
\begin{eqnarray*}
I_p(\rho;A) &\equiv& I_p(\rho;A,A) = \hbox{Tr}\left(\rho A^2\right) -\hbox{Tr}\left(  \rho^{\frac{1}{p}}A \rho^{\frac{1}{p^*}}A  \right)\\
 &=&-\frac{1}{2} \hbox{Tr}\left( [\rho^{\frac{1}{p}},A][\rho^{\frac{1}{p^*}},A]\right).
\end{eqnarray*}
We use the parameters $p$ and $p^*$, since many papers \cite{ig3,ig4,ig2,ig1} in this field use such notations.  
The Wigner-Yanase skew information is
\begin{eqnarray*}
I(\rho;A) &\equiv& I_2(\rho;A) = \hbox{Tr}\left(\rho A^2\right) - \hbox{Tr}\left(\rho^{\frac{1}{2}}A \rho^{\frac{1}{2}}A  \right) \\
&=&-\frac{1}{2} \hbox{Tr}\left( [\rho^{\frac{1}{2}},A]^2\right).
\end{eqnarray*}
\end{Def}

An interpretation of skew information as a measure of quantum uncertainty is
given in \cite{LZ} by S.Luo and Q.Zhang. They claimed the following uncertainty relation :
\begin{eqnarray} 
&&I(\rho ,A) I(\rho ,B) -\vert \hbox{Re} \left(\hbox{Corr}_{\rho}(A,B)\right) \vert^2 \nonumber \\
&&\geq \frac{1}{4} \vert \hbox{Tr}\left(\rho [A,B]\right)\vert^2, \label{error}
\end{eqnarray}
for two selfadjoint matrices $A$ and $B$, and density matrix $\rho$, where their correlation measure was defined by
$$\hbox{Corr}_{\rho}(A,B) \equiv \hbox{Tr}\left(\rho AB\right) - \hbox{Tr}\left(\rho^{1/2}A\rho^{1/2}B\right).$$
However, we show the inequality (\ref{error}) does not hold in general. We give a counter-example for inequality (\ref{error}) in the final section.

We define the generalized skew correlation and the generalized skew information as follows.

\begin{Def}
For arbitrary $X$ and $Y$, $p \in [1,+\infty]$ with $p^*$ such that $\frac{1}{p} + \frac{1}{p^*} = 1$  and $\varepsilon \geq 0$, set 
\begin{eqnarray*}
\phi_{p,\varepsilon}(\rho ; X,Y) &\equiv&  \varepsilon \hbox{Cov}_{\rho}(X^*,Y)\\
&+& \frac{1}{2}I_p(\rho;\widetilde{X^*},\widetilde{Y})+\frac{1}{2}I_p(\rho;\widetilde{Y},\widetilde{X^*}). 
\end{eqnarray*}
If $A$ and $B$ are selfadjoint, the generalized skew correlation is defined by
$$
\hbox{Corr}_{p,\varepsilon } \left(\rho; {A,B} \right) \equiv \phi_{p,\varepsilon}(\rho;A,B).
$$
The generalized skew information is defined by
$$
I_{p,\varepsilon } \left(\rho; A \right) \equiv \hbox{Corr}_{p,\varepsilon}(\rho;A,A) = \varepsilon V_{\rho}(A) + I_p(\rho;\widetilde{A}) $$
so that 
$$
I_{p,0}(\rho;A)=I_p(\rho;\widetilde{A}) = V_{\rho}(A)-\hbox{Tr}\left(  \rho^{\frac{1}{p}}\widetilde{A} \rho^{\frac{1}{p^*}}\widetilde{A}       \right).
$$
\end{Def}

Then we have the following theorem.

\begin{The}  \label{g_ur}
For any two selfadjoint matrices $A$ and $B$, any density matrix $\rho$, 
any $p \in [1,+\infty]$ with $p^*$ such that $\frac{1}{p} + \frac{1}{p^*} = 1$ and $\varepsilon \geq 0$, we have a generalized 
uncertainty relation :
\begin{eqnarray*}
&& I_{p,\varepsilon } \left(\rho; A \right)I_{p,\varepsilon } \left(\rho; B \right)
 - \left| {{\mathop{\rm Re}\nolimits} \left( {   \hbox{Corr}_{p,\varepsilon } \left(\rho; {A,B} \right)} \right)} \right|^2\\
&&  \ge \frac{{\varepsilon ^2 }}{4}\left| {\hbox{Tr}\left( {\rho \left[ {A,B} \right]} \right)} \right|^2 .
\end{eqnarray*}
\end{The}
{\it Proof} :
By Lemma \ref{g_Fujii}, $\phi_{p,\varepsilon}(\rho;X,X) \geq 0$. Furthermore it is clear that 
$\phi_{p,\varepsilon}(\rho;X,Y)  $ is sesquilinear and Hermitian. Then we have 
$$
|\phi_{p,\varepsilon}(\rho;X,Y)|^2 \leq \phi_{p,\varepsilon}(\rho;X,X) \phi_{p,\varepsilon}(\rho;Y,Y)
$$
by the Schwarz inequality.  It follows that 
$$
|\hbox{Corr}_{p,\varepsilon}(\rho;A,B)|^2 \leq \hbox{Corr}_{p,\varepsilon}(\rho;A,A) \hbox{Corr}_{p,\varepsilon}(\rho;B,B)
$$
for any two selfadjoint matrices $A$ and $B$. Then 
\begin{equation}  \label{schwarz}
|\hbox{Corr}_{p,\varepsilon}(\rho;A,B)|^2 \leq I_{p,\varepsilon}(\rho;A) I_{p,\varepsilon}(\rho;B)
\end{equation}
Simple calculations imply
\begin{eqnarray}
&& \hbox{Corr}_{p,\varepsilon } \left(\rho; {A,B} \right) - \hbox{Corr}_{p,\varepsilon } \left(\rho; {B,A} \right) \nonumber \\
&& = \varepsilon \hbox{Tr}\left( {\rho \left[ {\widetilde{A},\widetilde{B}} \right]} \right)  = \varepsilon \hbox{Tr}\left( {\rho \left[ {A,B} \right]} \right)   , \label{eq-1} \\
&& \hbox{Corr}_{p,\varepsilon } \left(\rho; {A,B} \right) + \hbox{Corr}_{p,\varepsilon } \left(\rho; {B,A} \right) \nonumber \\
&& = 2{\mathop{\rm Re}\nolimits} \left( {\hbox{Corr}_{p,\varepsilon } \left(\rho; {A,B} \right)} \right). \label{eq-2}
\end{eqnarray}
Summing both sides in the above two equalities, we have
\begin{eqnarray}
2\hbox{Corr}_{p,\varepsilon } \left(\rho; {A,B} \right) &=& \varepsilon \hbox{Tr}\left( {\rho \left[ {A,B} \right]} \right) \nonumber \\
&& +2{\mathop{\rm Re}\nolimits} \left( {\hbox{Corr}_{p,\varepsilon } \left(\rho; {A,B} \right)} \right).
\end{eqnarray}
Since $[A,B]$ is skew-adjoint, $\hbox{Tr}\left( {\rho \left[ {A,B} \right]} \right)$ is a purely imaginary number, we have
\begin{eqnarray}   
\left| {\hbox{Corr}_{p,\varepsilon } \left(\rho; {A,B} \right)} \right|^2 &=& 
\frac{{\varepsilon ^2 }}{4}\left| {\hbox{Tr}\left( {\rho \left[ {A,B} \right]} \right)} \right|^2 \nonumber \\ 
&+& \left| {{\mathop{\rm Re}\nolimits} \left( {\hbox{Corr}_{p,\varepsilon } \left(\rho; {A,B} \right)} \right)} \right|^2.  \label{eq1}
\end{eqnarray}
Thus the proof of the theorem is completed by the use of inequality (\ref{schwarz}) and Eq.(\ref{eq1}).

\hfill \qed

We are interested in the relationship between the left hand sides in Proposition \ref{ur} and Theorem \ref{g_ur}.
The following proposition gives the relationship.
\begin{Prop}   \label{order1}
For any two selfadjoint matrices $A$ and $B$, any density matrix $\rho$, any $p \in [1,+\infty]$ with $p^*$ such that $\frac{1}{p} + \frac{1}{p^*} = 1$ and $\varepsilon \geq 0$, we have
\begin{eqnarray*}
&&I_{p,\varepsilon } \left(\rho; A \right)I_{p,\varepsilon } \left( \rho;B \right) - 
\left| {{\mathop{\rm Re}\nolimits} \left( {\hbox{Corr}_{p,\varepsilon } \left(\rho; {A,B} \right)} \right)} \right|^2 \\
&& \ge \varepsilon ^2 V_\rho  \left( A \right)V_\rho  \left( B \right) - \varepsilon ^2 \left| {{\mathop{\rm Re}\nolimits} \left( {\hbox{Cov}_\rho  \left( {A,B} \right)} \right)} \right|^2. 
\end{eqnarray*}
\end{Prop}

{\it Proof} :
From Proposition \ref{ur}, we have $V_{\rho}(A)V_{\rho}(B) \geq \vert \hbox{Re} \left( \hbox{Cov}_{\rho}(A,B)\right) \vert^2 $, that is,
\begin{equation}  \label{eq2}
\left| {{   \mathop{\rm Re   }   \nolimits} \left( {\hbox{Tr}\left( {\rho \widetilde{A}\widetilde{B}} \right)} \right)} \right|^2
  \le \hbox{Tr}\left( {\rho \widetilde{A}^2 } \right)\hbox{Tr}\left( {\rho \widetilde{B}^2 } \right).
\end{equation}
By putting $\varepsilon = 0$ in (\ref{schwarz}), we have 
$$
|\hbox{Corr}_{p,0}(\rho;A,B)|^2 \leq I_{p,0}(\rho;A) I_{p,0}(\rho;B).
$$
It follows from (\ref{eq-1}) and (\ref{eq-2}) that 
$$
\hbox{Corr}_{p,0}(\rho;A,B) = {   \mathop{\rm Re   }   \nolimits} \left( \hbox{Corr}_{p,0}(\rho;A,B) \right).
$$
Thus 
\begin{equation}  \label{eq3}
|{   \mathop{\rm Re   }   \nolimits} \left( \hbox{Corr}_{p,0}(\rho;A,B) \right)|^2 \leq I_{p,0}(\rho;A) I_{p,0}(\rho;B).
\end{equation}
Using Eq.(\ref{eq2}), Eq.(\ref{eq3}) and direct calculations, we get the following:
\begin{eqnarray*}
&& L.H.S.-R.H.S. \\
&& = \varepsilon \hbox{Tr} \left(\rho \widetilde{A}^2\right)I_{p,0}(\rho;B)+ \varepsilon \hbox{Tr} \left(\rho \widetilde{B}^2\right) I_{p,0}(\rho;A) \\
&&-2 \varepsilon \hbox{Re} \left(\hbox{Tr} \left(\rho \widetilde{A}\widetilde{B}  \right)\right) \hbox{Re} \left(\hbox{Corr}_{p,0}(\rho;A,B)\right) \\
&& +I_{p,0}(\rho;A) I_{p,0}(\rho;B) - \left\{\hbox{Re} \left(\hbox{Corr}_{p,0}(\rho;A,B)\right) \right\}^2\\
&& \geq \varepsilon \hbox{Tr} \left(\rho \widetilde{A}^2\right)I_{p,0}(\rho;B)+ \varepsilon \hbox{Tr} \left(\rho \widetilde{B}^2\right) I_{p,0}(\rho;A) \\
&& -2 \varepsilon \hbox{Re} \left(\hbox{Tr} \left(\rho \widetilde{A}\widetilde{B}  \right)\right) \hbox{Re} \left(\hbox{Corr}_{p,0}(\rho;A,B)\right) \\
&& \geq \varepsilon \hbox{Tr} \left(\rho \widetilde{A}^2\right)I_{p,0}(\rho;B)+ \varepsilon \hbox{Tr} \left(\rho \widetilde{B}^2\right) I_{p,0}(\rho;A) \\
&& -2\varepsilon \sqrt{\hbox{Tr} \left(\rho \widetilde{A}^2\right)\hbox{Tr} \left(\rho \widetilde{B}^2\right) }\sqrt{I_{p,0}(\rho;A) I_{p,0}(\rho;B) }\\
&& = \varepsilon \left\{ \sqrt{\hbox{Tr} \left(\rho \widetilde{A}^2\right)I_{p,0}(\rho;B) } -\sqrt{\hbox{Tr} \left(\rho \widetilde{B}^2\right)I_{p,0}(\rho;A)}\right\}^2\\
&& \geq 0.
\end{eqnarray*}
\hfill \qed

\begin{Rem}
Theorem \ref{g_ur} can be also proven by Proposition \ref{ur} and Proposition \ref{order1}.
\end{Rem}

%%%%%%%%%%%%%%%%%%%%%%%%%%%%%%%%%%% Section %%%%%%%%%%%%%%%%%%%%%%%%%%%%%%%%%%%%%%%%%%%%%%%%%%%%%%%%%%%
\section{An inequality related to the uncertainty relation}
The trace inequality 
\begin{eqnarray*}
&& V_\rho  \left( A \right)V_\rho  \left( B \right) - \left| {{\mathop{\rm Re}\nolimits} \left( {\hbox{Cov}_\rho  \left( {A,B} \right)} \right)} \right|^2 \\
&& \ge I_{2,0} \left(\rho; A \right)I_{2,0} \left(\rho; B \right) - \left| {{\mathop{\rm Re}\nolimits} \left( {\hbox{Corr}_{2,0} \left(\rho; {A,B} \right)} \right)} \right|^2. 
\end{eqnarray*}
was conjectured in \cite{LZ2} and proven in \cite{LZ}. As a generalization of Theorem 2 in \cite{LZ}, we prove a one-parameter extention of the above inequality.
\begin{Prop}   \label{order2}
For any two selfadjoint matrices $A$ and $B$, any density matrix $\rho$ and any $p \in [1,+\infty]$ with $p^*$ such that $\frac{1}{p} + \frac{1}{p^*} = 1$, we have 
\begin{eqnarray}
\hspace*{-8mm}  && V_\rho  \left( A \right)V_\rho  \left( B \right) - \left| {{\mathop{\rm Re}\nolimits} \left( {\hbox{Cov}_\rho  \left( {A,B} \right)} \right)} \right|^2 \nonumber \\
\hspace*{-8mm}  && \ge I_{p,0} \left(\rho; A \right)I_{p,0} \left(\rho; B \right) - \left| {{\mathop{\rm Re}\nolimits} \left( {\hbox{Corr}_{p,0} \left(\rho; {A,B} \right)} \right)} \right|^2. \label{Kosaki_ineq} 
\end{eqnarray}
\end{Prop}
{\it Proof} :
Let $\left\{\varphi_i\right\}$ be a complete orthonormal basis composed by eigenvectors of $\rho$.
Then we calculate 
\[
\hbox{Tr} \left( {\rho ^{\frac{1}{p}}  \widetilde{A}\rho ^{\frac{1}{p^*}} \widetilde{A}} \right) = \sum\limits_{i,j} {\lambda _i^{\frac{1}{p}}  \lambda _j^{\frac{1}{p^*}}  a_{ij} a_{ji} }, 
\]
where $a_{ij}  \equiv \left\langle {\widetilde{A}\varphi _i \left| {\varphi _j } \right\rangle } \right.$ and $a_{ji}  \equiv \overline {a_{ij} } $.
Thus we get
\begin{eqnarray*}
&& I_{p,0} \left( \rho;  A \right) = V_\rho  \left( A \right) - \sum\limits_{i,j} {\lambda _i^{\frac{1}{p}}  \lambda _j^{\frac{1}{p^*}} a_{ij} a_{ji} } ,\\
&& I_{p,0} \left(\rho;   B \right) = V_\rho  \left( B \right) - \sum\limits_{i,j} {\lambda _i^{\frac{1}{p}}  \lambda _j^{\frac{1}{p^*}} b_{ij} b_{ji} },
\end{eqnarray*}
where $b_{ij}  \equiv \left\langle {\widetilde{B}\varphi _i \left| {\varphi _j } \right\rangle } \right.$ and $b_{ji}  \equiv \overline {b_{ij} }$.
In a similar way, we obtain
\begin{eqnarray*}
&& {\mathop{\rm Re}\nolimits} \left( {\hbox{Corr}_{p,0} \left(\rho;   {A,B} \right)} \right) = {\mathop{\rm Re}\nolimits} \left( {\hbox{Cov}_\rho  \left( {A,B} \right)} \right) \\
&& - \frac{1}{2}\sum\limits_{i,j} {\lambda _i^{\frac{1}{p}}  \lambda _j^{\frac{1}{p^*}} {\mathop{\rm Re}\nolimits} \left( {a_{ij} b_{ji} } \right)}  
- \frac{1}{2}\sum\limits_{j,i} {\lambda _i^{\frac{1}{p}}  \lambda _j^{\frac{1}{p^*}} {\mathop{\rm Re}\nolimits} \left( {b_{ij} a_{ji} } \right)}. 
\end{eqnarray*}
In order to prove the present proposition, we have only to show the inequality $\xi  \ge \eta$, where,
\begin{eqnarray}
&&\xi \equiv V_\rho  \left( A \right)\sum\limits_{i,j} {\lambda _i^{\frac{1}{p}}  \lambda _j^{\frac{1}{p^*}} b_{ij} b_{ji} }  + V_\rho  \left( B \right)\sum\limits_{i,j} {\lambda _i^{\frac{1}{p}}  \lambda _j^{\frac{1}{p^*}} a_{ij} a_{ji} } \nonumber \\
&&-\left( {\sum\limits_{i,j} {\lambda _i^{\frac{1}{p}}  \lambda _j^{\frac{1}{p^*}} a_{ij} a_{ji} } } \right)\left( {\sum\limits_{i,j} {\lambda _i^{\frac{1}{p}}  \lambda _j^{\frac{1}{p^*}} b_{ij} b_{ji} } } \right),\nonumber \\
&& \eta  \equiv {\mathop{\rm Re}\nolimits} \left( \hbox{Cov}_\rho  \left( {A,B} \right) \right)
\sum\limits_{i,j} \lambda _i^{\frac{1}{p}}  \lambda _j^{\frac{1}{p^*}} {\mathop{\rm Re}\nolimits} \left(a_{ij} b_{ji}\right)  \nonumber \\
&&+{\mathop{\rm Re}\nolimits} \left( \hbox{Cov}_\rho  \left( {A,B} \right)\right)
 \sum\limits_{i,j} \lambda _i^{\frac{1}{p}}  \lambda _j^{\frac{1}{p^*}} {\mathop{\rm Re}\nolimits} \left( b_{ij} a_{ji}\right)  \nonumber \\ 
&&-\frac{1}{4}\left( \sum\limits_{i,j} \lambda _i^{\frac{1}{p}}  \lambda _j^{\frac{1}{p^*}} {\mathop{\rm Re}\nolimits}  \left(a_{ij} b_{ji} \right)
 + \sum\limits_{i,j} \lambda _i^{\frac{1}{p}}  \lambda _j^{\frac{1}{p^*}} {\mathop{\rm Re}\nolimits} \left( b_{ij} a_{ji} \right)\right)^2. \nonumber 
\end{eqnarray}
Since $V_\rho  \left( A \right) = \hbox{Tr} \left( {\rho \widetilde{A}^2 } \right) = \frac{1}{2}\sum\limits_{i,j} {\left( {\lambda _i  + \lambda _j } \right)a_{ij} a_{ji} } ,
V_\rho  \left( B \right) = \hbox{Tr} \left( {\rho \widetilde{B}^2 } \right) = \frac{1}{2}\sum\limits_{i,j} {\left( {\lambda _i  + \lambda _j } \right)b_{ij} b_{ji} } $,
and $\left( {\lambda _i  + \lambda _j } \right)\lambda _k^{\frac{1}{p}}  \lambda _l^{\frac{1}{p^*}}  + \left( {\lambda _k  + \lambda _l } \right)\lambda _i^{\frac{1}{p}}  \lambda _j^{\frac{1}{p^*}}  - 2\lambda _i^{\frac{1}{p}}  \lambda _j^{\frac{1}{p^*}} \lambda _k^{\frac{1}{p}}  \lambda _l^{\frac{1}{p^*}}  \ge 0$,
we calculate
\begin{eqnarray}
&& \xi  = \frac{1}{4}\sum\limits_{i,j,k,l} \left\{ \left( \lambda _i  + \lambda _j  \right)\lambda _k^{\frac{1}{p}}  \lambda _l^{\frac{1}{p^*}}  + \left( \lambda _k  + \lambda _l  \right)\lambda _i^{\frac{1}{p}}  \lambda _j^{\frac{1}{p^*}}  \right. \nonumber \\
&& \left. - 2\lambda _i^{\frac{1}{p}}  \lambda _j^{\frac{1}{p^*}} \lambda _k^{\frac{1}{p}}  \lambda _l^{\frac{1}{p^*}}  \right\} \left( a_{ij} a_{ji} b_{kl} b_{lk}  + b_{ij} b_{ji} a_{kl} a_{lk}  \right) \nonumber \\ 
&&  \ge \frac{1}{2}\sum\limits_{i,j,k,l} \left\{ \left( {\lambda _i  + \lambda _j } \right)\lambda _k^{\frac{1}{p}}  \lambda _l^{\frac{1}{p^*}}  + \left( \lambda _k  + \lambda _l  \right)\lambda _i^{\frac{1}{p}}  \lambda _j^{\frac{1}{p^*}}   \right. \nonumber \\
&& \left. - 2\lambda _i^{\frac{1}{p}}  \lambda _j^{\frac{1}{p^*}} \lambda _k^{\frac{1}{p}}  \lambda _l^{\frac{1}{p^*}}  \right\}\left| a_{ij} b_{ji}  \right| \left| a_{kl} b_{lk}  \right| .
\end{eqnarray}
Since ${\mathop{\rm Re}\nolimits} \left( {b_{kl} a_{lk} } \right) = {\mathop{\rm Re}\nolimits} \left( {\overline {b_{lk} } \overline {a_{kl} } } \right)
 = {\mathop{\rm Re}\nolimits} \left( {b_{lk} a_{kl} } \right) = {\mathop{\rm Re}\nolimits} \left( {a_{kl} b_{lk} } \right),
{\mathop{\rm Re}\nolimits} \left( {b_{ij} a_{ji} } \right) = {\mathop{\rm Re}\nolimits} \left( {a_{ij} b_{ji} } \right)$,
we calculate
\begin{eqnarray*}
&& \eta = \frac{1}{2}\sum\limits_{i,j,k,l} \left\{ \left( \lambda _i  + \lambda _j  \right)\lambda _k^{\frac{1}{p}}  \lambda _l^{\frac{1}{p^*}}  + \left( \lambda _k  + \lambda _l  \right)\lambda _i^{\frac{1}{p}}  \lambda _j^{\frac{1}{p^*}}  \right. \\
&&\left. -2\lambda _i^{\frac{1}{p}}  \lambda _j^{\frac{1}{p^*}} \lambda _k^{\frac{1}{p}}  \lambda _l^{\frac{1}{p^*}}  \right\} 
{\mathop{\rm Re}\nolimits} \left( {a_{ij} b_{ji} } \right){\mathop{\rm Re}\nolimits} \left( {a_{kl} b_{lk} } \right).
\end{eqnarray*}
Thus we conclude $\xi  \geq  \eta$, since $
\left| {a_{ij} b_{ji} } \right|\left| {a_{kl} b_{lk} } \right| \ge \left| {{\mathop{\rm Re}\nolimits} \left( {a_{ij} b_{ji} } \right)
{\mathop{\rm Re}\nolimits} \left( {a_{kl} b_{lk} } \right)} \right|$.

\hfill \qed
%\begin{Rem}
%Proposition \ref{order2} is a kind of generalization of Theorem 2 in \cite{LZ}.
%\end{Rem}

The inequality (\ref{Kosaki_ineq}) was independently proven by H.Kosaki in \cite{Ko}. Our proof is simpler than Kosaki's one.

%%%%%%%%%%%%%%%%%%%%%%%%%%%%%%%%%%% Section %%%%%%%%%%%%%%%%%%%%%%%%%%%%%%%%%%%%%%%%%%%%%%%%%%%%%%%%%%%
%\section{Concluding remarks}
%We obtained a generalized uncertainty relation by introducing the generalized skew correlation and the generalized skew information.
%In the case of $\varepsilon = 0$ in Theorem \ref{g_ur}, we have physically meaningless relation which is represented by 
%the inequality Eq.(\ref{schwarz}) with $\varepsilon = 0$,
% since $\hbox{Corr}_{p,0}(\rho;A,B) = \hbox{Re} \left\{ \hbox{Corr}_{p,0} (\rho;A,B)\right\}$.
%On the other side, in the case of $\varepsilon \neq 0$, Proposition \ref{order1} gives the upper bound of the left hand side of the Schr\"odinger's uncertainty relation given in Eq.(\ref{eq_ur}).
As a concluding remark, we point out that Theorem 1 in \cite{LZ} is incorrect in general. 
\begin{Rem}
Theorem 1 in \cite{LZ} is not true in general. A counter-example is given as follows. Let 
\[
\rho  = \frac{1}{4}\left( \begin{array}{l}
 3\,\,\,\,\,0 \\ 
 0\,\,\,\,\,1 \\ 
 \end{array} \right),A = \left( \begin{array}{l}
 \,\,0\,\,\,\,\,\,\,\,i \\ 
  - i\,\,\,\,\,\,\,0 \\ 
 \end{array} \right),B = \left( \begin{array}{l}
 0\,\,\,\,\,1 \\ 
 1\,\,\,\,\,0 \\ 
 \end{array} \right).
\]
Then we have,
$I\left( {\rho ,A} \right)I\left( {\rho ,B} \right) - \left| {{\mathop{\rm Re}\nolimits} 
\left( {\hbox{Corr}_\rho  \left( {A,B} \right)} \right)} \right|^2  = \frac{{7 - 4\sqrt 3 }}{4}$ 
and $\left| {\hbox{Tr}\left( {\rho \left[ {A,B} \right]} \right)} \right|^2  = 1.$
These imply 
\[
I\left( {\rho ,A} \right)I\left( {\rho ,B} \right) - \left| {{\mathop{\rm Re}\nolimits} 
\left( {\hbox{Corr}_\rho  \left( {A,B} \right)} \right)} \right|^2 
 < \frac{1}{4}\left| {\hbox{Tr}\left( {\rho \left[ {A,B} \right]} \right)} \right|^2 .
\]
%Thus Theorem 1 in \cite{LZ} does not hold. Their mistake is caused by the use of the Schwarz inequality for their skew correlation defined in form Eq.(\ref{their_corr}) 
%although it does not have the nonnegativity for {\it any} linear operator. 
%Actually their skew correlation can be applied the inequality
%$$\vert \hbox{Re}\left\{ \hbox{Corr}_{\rho}(A,B) \right\}\vert ^2 \leq \hbox{Corr}_{\rho}(A,A) \hbox{Corr}_{\rho}(B,B),$$ 
%since it has the nonnegativity for {\it self-adjoint} operator. 
%However this application does not lead us to obtain a physically meaningful inequality such as an uncertainty relation.
\end{Rem}

\section*{Acknowledgment}
We would like to thank the reviewers for providing valuable comments to improve our manuscript.
%\section*{Acknowledgement}
%The authours thank referees for valuable comments on this paper.

\begin{biography}{Kenjiro YANAGI}
Kenjiro Yanagi(M'85) was born in Yamaguchi Prefecture, Japan, on October 8,
1950. He received the B.Sc., M.Sc., and D.Sc. degrees, all in information
sciences, from Tokyo Institute of Technology, Tokyo, Japan, in 1974, 1976,
and 1983, respectively. He was an Assistant Professor, a Lectureer, and
an Associate Professor in the Department of Mathematics, Faculty of Science,
Yamaguchi University, Yamaguchi, Japan, from 1976 to 1987, 1987 to 1989, and
1989 to 1993, respectively. Since 1993 he has been a Professor in the
Department of Applied Science, Faculty of Engineering, Yamaguchi University.
During 1984 -1985 he was on leave As a Researcher in the Department of
Statistics, University of North Carolina, Chapel Hill. His research interests
include mathematical information theory, applied functional Analysis,
measure theory, and fuzzy measure. Recent interests are quantum information
theory.
\end{biography}

\begin{biography}{Shigeru FURUICHI}
Shigeru Furuichi(M'98) received the B.S. degree in 1995, from Department of Mathematics, Tokyo University of Science, and M.S. and Ph.D. from Department of information Science,
  Tokyo University of Science, Japan, in 1997 and 2000, respectively. He was an Assistant Professor during 1997-2001 and has been a Lecturer since 2001 
in Department of Electronics and Computer Science,
Tokyo University of Science in Yamaguchi, Japan. His research interests are information theory, entropy theory and operator theory 
including quantum information theory as one of applications.
\end{biography}

\begin{biography}{Ken KURIYAMA}
Ken Kuriyama was born in Saga Prefecture, Japan, on March 18, 1947.
He received B.S. degree in 1971 in physics from Kyushu University, and
 M.S. and Ph.D. degrees in mathematics from Kyushu University, Fukuoka, Japan, in
1973 and 1981, respectively.
In 1977 he joined the Faculty of Yamaguchi University, where he is a Professor.
His current research include operator algebras in Hilbert spaces, quantum information
theory, mathematical programming and numerical analysis on rock mechanics.
\end{biography}

\end{document}